\documentstyle[12pt]{article}

\def\thebibliography#1{\bigskip\section*{\centering
References\\}\bigskip\list
  {}{\settowidth\labelwidth{#1}\leftmargin\labelwidth
    \advance\leftmargin\labelsep
    \usecounter{enumi}}
    \def\newblock{\hskip .11em plus .33em minus .07em}
    \sloppy\clubpenalty4000\widowpenalty4000
    \sfcode`\.=1000\relax}

\def\op#1{\mathop{\fam0 #1}\limits}
\newcommand{\nm}[1]{\mid {#1}\mid}
\newcommand{\id}{{\rm Id\,}}
\newcommand{\pr}{{\rm pr}}

\newcommand{\beq}{\begin{equation}}
\newcommand{\eeq}{\end{equation}}
\newcommand{\ben}{\begin{eqnarray}}
\newcommand{\een}{\end{eqnarray}}
\newcommand{\be}{\begin{eqnarray*}}
\newcommand{\ee}{\end{eqnarray*}}
\newcommand{\bea}{\begin{eqalph}}
\newcommand{\eea}{\end{eqalph}}

\newcommand{\bL}{{\bf L}}

\newcommand{\al}{\alpha}

\newcommand{\bt}{\beta}
\newcommand{\dl}{\delta}
\newcommand{\la}{\lambda}

\newcommand{\f}{\phi}

\newcommand{\om}{\omega}

\newcommand{\m}{\mu}
\newcommand{\n}{\nu}
\newcommand{\g}{\gamma}

\newcommand{\e}{\epsilon}

\newcommand{\vt}{\vartheta}

\newcommand{\si}{\sigma}
\newcommand{\Si}{\Sigma}

\newcommand{\wt}{\widetilde}
\newcommand{\wh}{\widehat}

\newcommand{\dr}{\partial}

\newcommand{\ar}{\op\longrightarrow}

\newcommand{\ot}{\otimes}

\let\ssection=\section
\renewcommand{\section}{\setcounter{equation}{0}\ssection}

\newcounter{eqalph}[section]
\newcounter{equationa}[section]
\newcounter{theorem}
\newcounter{proposition}
\newcounter{definition}

\def\thedefinition{\arabic{definition}}

\newenvironment{theo}{\refstepcounter{definition} \bigskip\noindent{\sc
Theorem \thedefinition}.}{\bigskip }
\newenvironment{prop}{\refstepcounter{definition} \bigskip\noindent{\sc
Proposition \thedefinition}.}{ \bigskip }

\newenvironment{eqalph}{\stepcounter{equation}
\setcounter{equationa}{\value{equation}}
\setcounter{equation}{0}

\begin{eqnarray}}{\end{eqnarray}
\setcounter{equation}{\value{equationa}}}

\begin{document}
\hbox{}

\noindent
{\large \bf COVARIANT SPIN STRUCTURE}
\bigskip

\noindent
{\sc G. Sardanashvily} 
\medskip

\noindent
Department of Theoretical Physics, Moscow State University
117234 Moscow, Russia\newline
E-mail: sard@grav.phys.msu.su
\bigskip

{\small
\noindent
Every Dirac spin structure on a world manifold is associated with a
certain gravitational field, and is not preserved under general
covariant transformations. We construct a composite spinor bundle such
that any Dirac spin structure is its subbundle, and this bundle admits
general covariant transformations.}

\section{Introduction}

Metric and metric-affine theories of gravity
are formulated on the natural bundles
$Y\to X$ (e.g., tensor bundles) which admit the canonical
lifts of diffeomorphisms of the base $X$. These lifts are general covariant
transformations of
$Y$. The invariance of a gravitational Lagrangian under these
transformations leads to the energy-momentum conservation law, where the
gravitational energy-momentum flow reduces to the generalized Komar
superpotential 
(Novotn\'y, 1984; Borowiec {\it et al.}, 1994;  Giachetta and
Sardanashvily, 1996; Giachetta and Mangiarotti, 1997; Sardanashvily, 1997). 
Difficulties arise in gauge gravitation theory in the presence
of Dirac fermion
fields. The corresponding spin structure is associated with a certain
gravitational field, and it is not preserved under 
general covariant transformations. 

To overcome these difficulties, we will
consider the universal two-fold covering group $\wt{GL_4}$ of the general
linear group $GL_4=GL^+(4,{\bf R})$
and the corresponding two-fold covering bundle $\wt{LX}$ of the bundle of
linear frames $LX$ (Dabrowski and Percacci, 1986; Lawson and Michelson,
1989; Switt, 1993).  
 One can consider the spinor representations
of the group $\wt{GL}_4$ which, however, are infinite-dimensional
(Hehl {\it et al.}, 1995). 
At the same time, the following procedure enables us not to exceed the
scope of standard fermion models.
The total space of the $\wt{GL_4}$-principal bundle 
$\wt{LX}\to X$ is the $L_s$-principal bundle $\wt{LX}\to \Si_T$ 
with the structure group  $L_s=SL(2,{\bf C})$ over the
quotient bundle
\be
\Si_T=\wt{LX}/L_s \to X,
\ee
whose sections are tetrad gravitational fields $h$.
Let us consider the spinor bundle 
\be
S=(\wt{LX}\times V)/L_s
\ee
associated with the principal bundle $\wt{LX}\to \Si_T$. Given a
tetrad field $h$, the restriction of $S$ to $h(X)\subset \Si_T$
is a subbundle of the composite spinor bundle
\be
S\to \Si_T \to X
\ee
which is exactly the spin structure associated with the
gravitational field $h$. General covariant transformations 
of the frame
bundle $LX$ and, consequently, of the bundles $\wt{LX}$ 
and $S$ are defined.

\section{Preliminaries}

Manifolds throughout are real, finite-dimensional,
Hausdorff, second-countable and connected. By a world
manifold
$X$ is meant a 4-dimensional manifold which is non-compact,
orientable, and parallelizable in order that a
pseudo-Riemannian metric, a spin structure, and a 
causal space-time structure
to exist on $X$. Note that every non-compact manifold admits a 
pseudo-Riemannian metric, and a non-compact 4-dimensional 
manifold $X$ has a spin structure iff it
is parallelizable.
Moreover, this spin structure is unique (Geroch, 1968; Avis and Isham, 1980). 

Let $\pi_{LX}:LX\to X$
be the principal bundle of oriented linear
frames in the tangent spaces to a world manifold $X$ (or simply the  frame
bundle). Its structure group is 
$GL_4.$ A world manifold $X$, by definition, is parallelizable iff the
frame bundle
$LX\to X$ is trivial. 
Given the holonomic frames
$\{\dr_\m\}$ in the tangent bundle $TX$,
every element
$\{H_a\}$ of the frame bundle
$LX$ takes the form 
$H_a=H^\m{}_a\dr_\m$,
where $H^\m{}_a$ is a matrix element of the group $GL_4$. 
The frame bundle $LX$
is provided with the bundle coordinates 
$(x^\la, H^\m{}_a)$.
In these coordinates, the canonical action  
of the structure group $GL_4$ on $LX$ reads 
\be
R_g: H^\m{}_a\mapsto H^\m{}_bg^b{}_a, \qquad g\in GL_4.
\ee
The frame bundle $LX$ is equipped with the canonical ${\bf
R}^4$-valued 1-form 
\beq
\theta_{LX} = H^a{}_\m dx^\m\ot t_a,\label{b3133'}
\eeq
where $\{t_a\}$ is a fixed basis for ${\bf R}^4$ and $H^b{}_\m$ is the 
inverse
matrix of $H^\m{}_a$.

The frame bundle $LX\to X$ belongs to the category of natural bundles. 
Every diffeomorphism $f$ of $X$ gives rise canonically to the automorphism
\beq
\wt f: (x^\la, H^\la{}_a)\mapsto (f^\la(x),\dr_\m f^\la H^\m{}_a) \label{025}
\eeq
of $LX$ and to the corresponding
automorphisms (general covariant transformations)
\be
\wt f:T=(LX\times V)/GL_4\to (\wt f(LX)\times V)/GL_4
\ee
of any fibre bundle $T$ associated with $LX$.
In particular, if $T=TX$, the lift $\wt f=Tf$ 
is the familiar tangent morphism to $f$.

The lift (\ref{025}) yields the  canonical horizontal lift
 $\wt\tau$ of every vector
field $\tau$ on
$X$ onto the principal bundle $LX$ and the associated bundles. The
canonical lift of $\tau$ over $LX$ is defined by the relation
$\bL_{\wt\tau}\theta_{LX}=0$.
The corresponding
canonical lift 
of $\tau$ onto the tensor bundle 
\be
(\op\ot^mTX)\ot(\op\ot^kT^*X). 
\ee
reads
\beq
\wt\tau = \tau^\m\dr_\m + [\dr_\nu\tau^{\al_1}\dot
x^{\nu\al_2\cdots\al_m}_{\bt_1\cdots\bt_k} + \ldots
-\dr_{\bt_1}\tau^\nu \dot x^{\al_1\cdots\al_m}_{\nu\bt_2\cdots\bt_k}
-\ldots]\frac{\dr}{\dr \dot
x^{\al_1\cdots\al_m}_{\bt_1\cdots\bt_k}}. \label{l28}
\eeq

A  pseudo-Riemannian metric $g$
 on a world manifold
$X$, called a world metric, is represented by a section of the metric bundle 
\beq
\Si_{PR}= GLX/O(1,3),\label{b3203}
\eeq
where  by $GLX$ is meant the bundle of all
linear frames in $TX$ and $O(1,3)$ is the complete Lorentz group. 
Since $X$ is oriented, $\Si_{PR}$ is
associated with the bundle $LX$ of oriented frames in $TX$.
Its typical fibre
is the quotient space $GL(4,{\bf R})/O(1,3)$,
homeomorphic to the topological space
${\bf RP}^3\times
{\bf R}^7$, where by ${\bf RP}^3$ is meant the
3-dimensional real projective space. 
For the sake of siplicity, we will often identify the metric
bundle with an open subbundle of the tensor bundle
$\Si_{PR}\subset \op\vee^2 TX$
with coordinates $(x^\la, \si^{\m\nu})$. 
By $\si_{\m\n}$ are
meant the components of the inverse matrix, and  $\si =\det(\si_{\m\n})$.
The canonical lift $\wt\tau$ (\ref{l28}) onto $\Si_{PR}$ 
reads
\beq
\wt\tau =\tau^\la\dr_\la + 
(\dr_\nu\tau^\al\si^{\nu\bt}
+\dr_\nu\tau^\bt\si^{\nu\al})\frac{\dr}{\dr\si^{\al\bt}}.
\label{973}
\eeq

A linear connection on
$TX$ and $T^*X$, called a world connection, 
is given by coordinate expressions
\ben
&& K= dx^\la\otimes (\dr_\la +K_\la{}^\m{}_\n
\dot x^\n
\frac{\dr}{\dr\dot x^\m}), \label{08} \\
&& K^*= dx^\la\otimes (\dr_\la -K_\la{}^\m{}_\n \dot x_\m
\frac{\dr}{\dr\dot x_\n}). \label{08'}
\een
There is one-to-one correspondence
between the world connections and the sections of 
the quotient bundle 
\beq
C_K=J^1LX/GL_4, \label{015}
\eeq
where by $J^1LX$ is meant the first order jet manifold of the frame bundle
$LX\to X$. 
With respect to the holonomic frames in $TX$, the bundle $C_K$ 
is coordinatized by $(x^\la, k_\la{}^\nu{}_{\al})$ 
so that, for any section $K$
of $C_K\to X$,
\be
k_\la{}^\nu{}_\al\circ K=K_\la{}^\nu{}_\al
\ee
are the coefficients of the world connection $K$ (\ref{08}). 
There exists the canonical lift
\beq
\wt\tau = \tau^\m\dr_\m +[\dr_\nu\tau^\al k_\m{}^\nu{}_\bt -
\dr_\bt\tau^\nu k_\m{}^\al{}_\nu - \dr_\m\tau^\nu
k_\nu{}^\al{}_\bt + \dr_{\m\bt}\tau^\al]\frac{\dr}{\dr k_\m{}^\al{}_\bt}
\label{b3150}
\eeq
onto $C_K$ of a vector field $\tau$ on $X$.

Note that, if a vector field $\tau$ is non-vanishing 
at a point $x\in X$, then
there exists a local symmetric connection $K$ around $x$ such that
$\tau$ is its integral section, i.e.,
$\dr_\nu\tau^\al =K_\nu{}^\al{}_\bt\tau^\bt$.
Then the canonical lift $\wt\tau$ (\ref{l28}) can be found locally as the
horizontal lift of $\tau$ by $K$.

\section{Dirac spinors}

Let $M$ be the Minkowski space equipped with the Minkowski metric which
reads
\be
\eta ={\rm diag}(1,-1,-1,-1)
\ee
with respect to a basis $\{e^a\}$ for $M$.
Let ${\bf C}_{1,3}$ be the complex  Clifford
algebra  generated by elements of
$M$. 
It is isomorphic to the real
Clifford algebra ${\bf R}_{2,3}$,  whose generating
space is ${\bf R}^5$ with the metric 
${\rm diag}(1,-1,-1,-1,1).$ Its
subalgebra generated by the elements of $M\subset {\bf
R}^5$ is the real Clifford algebra 
${\bf R}_{1,3}$.

A  spinor space $V$ is defined to be a
minimal left ideal of ${\bf C}_{1,3}$ (Crawford, 1991; Rodrigues and De
Souza, 1993; Obukhov and
Solodukhin, 1994). 
We have the representation
\beq
\g: M\otimes V \to V, \qquad \g(e^a)=\g^a, \label{w01}
\eeq
of elements of the Minkowski space $M\subset{\bf C}_{1,3}$ by the Dirac
$\g$-matrices on $V$. Different 
ideals lead to equivalent
representations (\ref{w01}).
The spinor space $V$ is provided with 
the  spinor metric 
\beq
a(v,v') =\frac12(v^+\g^0v' +{v'}^+\g^0v),
\label{b3201}
\eeq
since the element $e^0\in M$ satisfies the
conditions
\be
(e^0)^+=e^0, \qquad (e^0e)^+=e^0e, \qquad \forall e\in M.
\ee

The  Clifford group $G_{1,3}$
comprises all invertible elements $l_s$ of the real Clifford
algebra ${\bf R}_{1,3}$ such that the corresponding inner automorphisms
preserve 
the Minkowski space $M\subset {\bf R}_{1,3}$, that is,
\beq
l_sel^{-1}_s = l(e), \qquad e\in M, \label{b3200}
\eeq
where $l\in O(1,3)$ is a Lorentz transformation of $M$. 
Thus, we have an
epimorphism of the Clifford group $G_{1,3}$ onto the Lorentz group
$O(1,3)$. Since the action (\ref{b3200}) of the Clifford group on the
Minkowski space 
$M$ is not effective, one usually considers its  pin
and spin subgroups.  
The 
subgroup $Pin(1,3)$ of $G_{1,3}$ is generated by elements $e\in
M$ such that $\eta(e,e)=\pm 1$.  
The even part of 
$Pin(1,3)$ is the spin group $Spin(1,3)$. 
Its component of the unity 
$L_s\simeq SL(2,{\bf C})$
is the two-fold universal covering group
$z_L:L_s\to L$
of the proper  
Lorentz group $L=SO^0(1,3)$. Recall
that $L$ is homeomorphic to ${\bf RP}^3\times {\bf R}^3$. 
The Lorentz group $L$ acts on the Minkowski space $M$ by the
generators
\beq
L_{ab}{}^c{}_d= \eta_{ad}\dl^c_b- \eta_{bd}\dl^c_a. \label{b3278}
\eeq

The Clifford group $G_{1,3}$ acts on the spinor space $V$ by left
multiplications 
\be
G_{1,3}\ni l_s:v\mapsto l_sv, \qquad v\in V.
\ee
This action preserves the
representation (\ref{w01}), i.e.,
\beq
\g (lM\otimes l_sV) = l_s\g (M\otimes V). \label{b3191}
\eeq
The spin group
$L_r$ acts on the spinor space $V$ by the generators
\beq
L_{ab}=\frac{1}{4}[\g_a,\g_b]. \label{b3213}
\eeq
Since $L_{ab}^+\g^0=- \g^0L_{ab}$,
this action preserves the spinor metric (\ref{b3201}).

Let us consider a bundle of 
Minkowski spaces $MX\to X$ over a world manifold $X$. This bundle
is extended to the bundle of Clifford algebras $CX$ with the fibres $C_xX$
generated by the fibres $M_xX$ of $MX$ (Benn and Tucker, 1987;
Rodrigues and Vaz, 1996). 
The bundle $CX$ has the structure 
group Aut$({\bf C}_{1,3})$ of inner 
automorphisms of the Clifford 
algebra ${\bf C}_{1,3}$. This structure group is 
reducible to the Lorentz group $SO(1,3)$, and the 
bundle of Clifford algebras $CX$ contains the 
subbundle
$MX$ of the generating Minkowski spaces. 
However, $CX$ does not necessarily
contain a spinor subbundle because a 
spinor subspace $V$ of ${\bf C}_{1,3}$
is not stable under inner automorphisms of ${\bf C}_{1,3}$. 
As was shown (Benn and Tucker, 1988;
Rodrigues and Vaz, 1996), the above-mentioned  
spinor subbundle $S_M$ exists if
the transition functions of $CX$ 
can be lifted from Aut$({\bf C}_{1,3})$ to $G_{1,3}$. This agrees with
the usual conditions of existence of a spin structure. 
The bundle
$MX$ of Minkowski spaces must be isomorphic to the cotangent bundle
$T^*X$ for sections of the spinor bundle $S_M$ to describe Dirac
fermion fields on a world manifold
$X$. In other words, we should consider a spin structure on the cotangent
bundle $T^*X$ of $X$ (Lawson and Michelson, 1989).
There are several almost equivalent definitions of such a spin structure
(Avis and Isham, 1980; 
Benn and Tucker, 1987; Lawson and Michelson, 1989; Van der Heuvel, 1994). 
A Dirac spin structure 
 on a world manifold $X$ is said to be a pair
$(P_s, z_s)$ of an $L_s$-principal bundle $P_s\to X$ and a principal
bundle morphism
\beq
z_s: P_s \to LX. \label{b3246}
\eeq
Since the homomorphism $L_s \to GL_4$
factorizes through the epimorphism $L_s\to L$, every bundle morphism
(\ref{b3246}) factorizes through a morphism of $P_s$ onto some
$L$-principal subbundle of the frame bundle $LX$. 
It follows that the necessary condition for the
existence of a Dirac spin structure on $X$ is that 
the structure group $GL_4$ of 
$LX$ is reducible to the proper Lorentz group $L$. 

\section{Reduced structure}

Let 
$\pi_{PX}:P\to X$
be a principal bundle with a structure group $G$, which acts freely and
transitively on $P$ on the right:
\beq
R_g : p\mapsto pg, \quad  p\in P,\quad g\in G. \label{b1}
\eeq
Let 
\beq
Y=(P\times V)/G \label{b3103}
\eeq
be a $P$-associated bundle with 
a typical fibre $V$ on which the
structure group $G$ acts on the left.  
By $[p]$ we denote
the restriction of the canonical morphism
\be
P\times V\to (P\times V)/G
\ee
to $\{p\}\times V$ and
write $[p](v)= (p,v)\cdot G$.

By a principal automorphism of a principal bundle $P$ is
meant its automorphism $\Phi$ which is
equivariant under the canonical action (\ref{b1}), that is, 
$R_g\circ \Phi=\Phi\circ R_g$ for all $g\in G$.
A principal automorphism yields the corresponding
automorphism 
\beq
\Phi_Y: (P\times V)/G\to  (\Phi(P)\times V)/G \label{024}
\eeq
of every $P$-associated bundle $Y$ (\ref{b3103}). An automorphism $\Phi$
over $\id_X$ is called vertical.

Let $H$ be a  
Lie subgroup of $G$. We have the composite bundle
\beq
P\to P/H\to X, \label{b3223a}
\eeq
where
\beq
\Si=P/H\ar^{\pi_{\Si X}} X \label{b3193}
\eeq
is a $P$-associated bundle with the typical fibre $G/H$, and 
\beq
P_\Si=P\ar^{\pi_{P\Si}} P/H \label{b3194}
\eeq
is a principal bundle with the structure group $H$.
The structure group $G$ of a principal bundle $P$ is said to be reducible
to the subgroup
$H$ if there exists a $H$-principal
subbundle
$P^h$ of $P$. This subbundle is called a
reduced $G^\downarrow H$-structure (Kobayashi, 1972; Gordejuela and
Masqu\'e, 1995).
Recall the following theorem (Kobayashi and Nomizu, 1963). 

\begin{theo}\label{redsub}
There is one-to-one correspondence $P^h=\pi_{P\Si}^{-1}(h(X))$
between the $H$-principal subbundles 
$P^h$ of $P$ and the global sections $h$ of the quotient bundle
$P/H\to X$. Given such a section $h$, let us consider 
the restriction $h^*P_\Si$ of the
$H$-principal bundle
$P_\Si$ (\ref{b3194}) to $h(X)$. This is an $H$-principal bundle over $X$,
which is isomorphic to the reduced
subbundle
$P^h$ of $P$.   
\end{theo}

In general, there are topological obstructions to the reduction of a
structure group of a principal 
bundle to a subgroup. A structure group $G$ of a
principal bundle $P$ is reducible to its closed subgroup $H$ if the
quotient
$G/H$ is homeomorphic to a Euclidean space. 
In this case, all
$H$-principal subbundles of $P$ are isomorphic to each
other as
$H$-principal bundles (Steenrod, 1972). 
In particular, a structure group
is always reducible
to its maximal compact subgroup.
The following assertions take place (Giachetta {\it et al.}, 1997).

\begin{prop}\label{isomorp1}
Every vertical principal automorphism $\Phi$ of the
principal bundle
$P\to X$ sends a reduced subbundle $P^h$ onto an isomorphic $H$-principal
subbundle
$P^{h'}$. 
Conversely, let two reduced subbundles $P^h$ and $P^{h'}$ of a principal
bundle $P$ be isomorphic to each other as $H$-principal bundles and
let $\Phi:P^h\to P^{h'}$ be an isomorphism. Then $\Phi$ can be extended to 
a vertical automorphism of $P$. 
\end{prop}

Given a reduced subbundle $P^h$ of a principal bundle $P$, let
\beq
Y^h=(P^h\times V)/H \label{int1}
\eeq
be the $P^h$-associated bundle with a typical fibre $V$.
If $P^{h'}$ is another reduced 
subbundle of $P$ which is isomorphic to $P^h$,
the fibre bundles $Y^h$ and $Y^{h'}$ are isomorphic, but not canonically
isomorphic in general.

\begin{prop}\label{iso1}  Let $P^h$ be an $H$-principal 
subbundle of a
$G$-principal bundle $P$. Let $Y$ be the $P^h$-associated bundle
(\ref{int1}) with a typical fibre $V$. If $V$
carries a representation of the whole group $G$, 
the fibre bundle $Y^h$ is
canonically isomorphic to the $P$-associated
fibre bundle (\ref{b3103}).
\end{prop}

It follows that, given a $H$-principal 
subbundle $P^h$ of $P$, any
$P$-associated bundle
$Y$ with the structure group $G$ is canonically equipped 
with a structure of the
$P^h$-associated fibre bundle $Y^h$ with the structure group $H$. 
Briefly, we will write
\be
Y=(P\times V)/G =(P^h\times V)/H.
\ee
However, $P^h$- and $P^{h'}$-associated bundle
structures on $Y$ are not equivalent because, given  
bundle atlases $\Psi^h$ of
$P^h$ and $\Psi^{h'}$ of $P^{h'}$, the union of the associated atlases
of $Y$ has necessarily $G$-valued transition 
functions between the charts from $\Psi^h$ and $\Psi^{h'}$.

\section{Dirac spin structure}

Since a world manifold is parallelizable, the structure group
$GL_4$ of the frame bundle $LX$ is reducible to the Lorentz group 
$L$.
The corresponding $L$-principal subbundle $L^hX$ 
is said to be a Lorentz structure.

In accordance with Theorem \ref{redsub}, there is
one-to-one correspondence between the $L$-principal subbundles
$L^hX$ of
$LX$ and the global
sections $h$ of the quotient bundle 
\beq
\Si_T=LX/L, \label{5.15}
\eeq
called the  tetrad bundle. 
This  is an $LX$-associated bundle with the typical 
fibre
$GL_4/L$. Since the group $GL_4$ is 
homotopic to its maximal compact subgroup
$SO(4)$ and the Lorentz group $L$ is homotopic to 
$SO(3)$, $GL_4/L$ is homotopic to 
$SO(4)/SO(3)=S^3$, and homeomorphic to 
$S^3\times{\bf R}^7$. The bundle (\ref{5.15}) is the two-fold covering of
the metric bundle
$\Si_{PR}$ (\ref{b3203}). Its global sections are called the tetrad
fields.

Since $X$ is parallelizable, any two Lorentz subbundles $L^hX$ and $L^{h'}X$
are isomorphic to each other. By virtue of Proposition
\ref{isomorp1}, there exists a vertical bundle automorphism
$\Phi$ of $LX$ which sends $L^hX$ onto $L^{h'}X$.
The associated  vertical automorphism $\Phi_\Si$ of the fibre bundle
$\Si_T\to X$ transforms the tetrad field $h$ into the tetrad field $h'$.

Every tetrad field $h$ defines an associated  Lorentz atlas
$\Psi^h=\{U_\zeta,z_\zeta^h \}$ of 
$LX$ such that the corresponding local sections $z_\zeta^h$ of the frame
bundle $LX$ take their values into the Lorentz subbundle $L^hX$. 
Given a Lorentz atlas $\Psi^h$, the
pull-back 
\beq
z_\zeta^{h*}\theta_{LX}=h^a\ot t_a=h_\la^a dx^\la\ot t_a \label{b3211}
\eeq
of the canonical form $\theta_{LX}$ (\ref{b3133'}) by a local
section
$z_\zeta^h$ is said to be a (local)  tetrad form. 
It determines the  tetrad
coframes
\be
t^a(x) = h^a_\m(x)dx^\m, \qquad x\in U_\zeta,
\ee
in the cotangent bundle $T^*X$, which are associated with the
Lorentz atlas $\Psi^h$.
The coefficients $h^a_\m$ of the 
tetrad forms and the inverse matrix elements
$h^\m_a =H^\m_a\circ z^h_\zeta$
are called tetrad functions.  
Given a Lorentz atlas $\Psi^h$, the tetrad 
field $h$ can be represented by the
family of tetrad functions $\{h^\m_a\}$.
We have the familiar
relation $g=h^a\ot h^b\eta_{ab}$ 
between tetrad and metric fields. 

Given a tetrad field $h$, let $L^hX$ be the corresponding Lorentz
subbundle. Since $X$ is non-compact and 
parallelizable, the principal bundle
$L^hX$ can be extended uniquely to a
$L_s$-principal bundle $P^h\to X$, 
called the $h$-associated principal spinor
bundle (Geroch, 1968). 
We have the principal bundle morphism 
\beq
z_h: P^h \to L^hX, 
\qquad z_h\circ R_g =R_{z_L(g)}, \qquad \forall g\in L_s. \label{b3195}
\eeq
This is the $h$-associated Dirac spin structure 
on a world manifold.

Let us consider the $L^hX$-associated bundle of
Minkowski spaces
\beq
M^hX=(L^hX\times M)/L=(P^h\times M)/L_s \label{b3192}
\eeq
and the $P^h$-associated spinor bundle
\beq
S^h=(P^h\times V)/L_s,\label{510}
\eeq
called the  $h$-associated
spinor bundle. By virtue of Proposition \ref{iso1},
the bundle 
$M^hX$ (\ref{b3192}) is isomorphic to the cotangent
bundle 
\beq
T^*X=(L^hX\times M)/L. \label{int2}
\eeq
Then there exists the representation
\beq
\g_h: T^*X\ot S^h=(P^h\times (M\ot V))/L_s\to (P^h\times
\g(M\ot V))/L_s=S^h \label{L4}
\eeq
of covectors to $X$ by the Dirac $\g$-matrices
on elements of the spinor bundle $S^h$. 
Relative to an atlas $\{z_\zeta\}$ of $P^h$ and to 
the associated Lorentz atlas
$\{z^h_\zeta=z_h\circ z_\zeta\}$ of $LX$, the
representation (\ref{L4}) reads
\be
y^A(\g_h(h^a(x) \ot v))=\g^{aA}{}_By^B(v), \qquad v\in S^h_x,
\ee
where $y^A$ are the
corresponding 
bundle coordinates of $S^h$, and $h^a$ are the tetrad coframes
(\ref{b3211}). 
For brevity, we will write
\be
\wh h^a=\g_h(h^a)=\g^a,\qquad 
\wh dx^\la=\g_h(dx^\la)=h^\la_a(x)\g^a.
\ee

Sections $s_h$ of the $h$-associated spinor 
bundle $S^h$ (\ref{510}) describe
Dirac fermion fields in the presence of the tetrad field $h$. Indeed,
let $A_h$ be a principal connection on $S^h$, and let
\be
&&D: J^1S^h\to T^*X\op\ot_{S^h} S^h,\\
&&D=(y^A_\la-A^{ab}{}_\la L_{ab}{}^A{}_By^B)dx^\la\ot\dr_A,
\ee
be the corresponding covariant differential.
The first order
differential  Dirac operator is defined on $S^h$
by the composition 
\ben
&& \Delta_h=\g_h\circ D: J^1S^h\to T^*X\ot S^h\to S^h, \label{l13}\\
&& y^A\circ\Delta_h=h^\la_a \g^{aA}{}_B(y^B_\la- \frac12 A^{ab}{}_\la
L_{ab}{}^A{}_By^B). \nonumber
\een

The $h$-associated spinor bundle $S^h$ is equipped with the  fibre spinor
metric 
\be
a_h(v,v')=\frac12(v^+\g^0v' +{v'}^+\g^0v), \qquad v,v'\in S^h.
\ee
Using this metric and the Dirac operator (\ref{l13}), one can define the
 Lagrangian density 
\ben
&& L_h=\{\frac{i}{2}h^\la_q[y^+_A(\g^0\g^q)^A{}_B(y^B_\la
-\frac12A_\la{}^{ab} L_{ab}{}^B{}_Cy^C) - \label{b3215}\\
&& \qquad (y^+_{\la A} -\frac12A_\la{}^{ab}y^+_CL_{ab}^+)
(\g^0\g^q)^A{}_By^B] - my^+_A(\g^0)^A{}_By^B\}\det(h^a_\m) \nonumber
\een
on $J^1S^h$ which describes Dirac fermion fields in the
presence of a background tetrad field $h$ and a background connection $A_h$
on $S^h$.

\section{Spin connections}

Let us recall the following theorem (Kobayashi and Nomizu, 1963). 

\begin{theo}\label{connmorp}
Let $P'\to X$ and $P\to X$ be principle bundles with the structure groups
$G'$ and $G$, respectively. Let $\Phi: P'\to P$ be a bundle morphism
over
$X$ with the corresponding homomorphism $G'\to G$. For every 
principal connection
$A'$ on $P'$, there exists a unique
principal connection
$A$ on $P$ such that $T\Phi$
sends the horizontal subspaces of $A'$ onto the horizontal subspaces of $A$ 
\end{theo}

It follows that every principal (spin) connection 
\beq
A_h=dx^\la\ot(\dr_\la + \frac12A_\la{}^{ab} e_{ab}) 
\label{b3205}
\eeq
on $P^h$ defines a principal (Lorentz) connection on
$L^hX$ which is given by the same expression (\ref{b3205}).
Conversely, the pull-back $z_h^*\om_A$ on
$P^h$ of a connection form $\om_A$ of a 
Lorentz connection $A_h$ on $L^hX$ is
equivariant under the action of group
$L_s$ on
$P^h$ and, consequently, it 
is a connection form of a spin connection on $P^h$. 
In particular, the Levi--Civita connection of a
tetrad field 
$h$ gives rise to a spin connection 
\beq
A_\la{}^{ab}=\eta^{kb}h^a_\m(\dr_\la h^\m_k - h^\nu_k\{_\la{}^\m{}_\nu\})
\label{b3217}
\eeq
on the $h$-associated spinor
bundle $S^h$.

Moreover, every world connection $K$ on a world manifold $X$ also
defines a spin connection on a $h$-associated principal spinor bundle
$P^h$ (Giachetta and Mangiarotti, 1997; Sardanashvily, 1997).

In accordance with Theorem \ref{connmorp}, every Lorentz
connection $A_h$ (\ref{b3205})
on a Lorentz subbundle $L^hX$ of $LX$ gives rise to a world
connection 
\beq
K_\la{}^\m{}_\nu = h^k_\nu\dr_\la h^\m_k + \eta_{ka}h^\m_b h^k_\nu
A_\la{}^{ab} \label{b3207}
\eeq
on $LX$.
At the same time, every principal connection $K$ on 
the frame bundle
$LX$ defines a Lorentz principal connection $K_h$ on 
a $L$-principal subbundle
$L^hX$ as follows.

It is readily observed that the Lie algebra of the general linear group
$GL_4$ is the direct sum
\be
{\bf g}(GL_4) = {\bf g}(L) \oplus {\bf m}
\ee
of the Lie algebra ${\bf g}(L)$ of the Lorentz group and a subspace ${\bf
m}\subset {\bf g}(GL_4)$ such that 
$ad(l)({\bf m})\subset {\bf m}$, for all $l\in L$.
Let $\om_K$ 
be a connection form of a
world connection $K$ on $LX$. Then, by the well-known theorem
(Kobayashi and Nomizu, 1963),
the pull-back over $L^hX$ of the ${\bf g}(L)$-valued 
component $\om_L$
of
$\om_K$ is a connection form of a principal connection $K_h$ on the 
Lorentz subbundle $L^hX$. To obtain $K_h$, 
let us consider a local connection 1-form of the connection $K$ with
respect to a Lorentz atlas $\Psi^h$ of $LX$ given by the tetrad forms
$h^a$. This reads
\be
&& {z^h}^*\om_K= K_\la{}^b{}_k dx^\la\ot e_b{}^k,\\
&& K_\la{}^b{}_k = -h^b_\m \dr_\la h^\m_k  + K_\la{}^\m{}_\nu h^b_\m
h^\nu_k,
\ee
where $\{e_b{}^k\}$ is the basis for the Lie algebra of the group $GL_4$.
Then, the Lorentz part of this form is presicely the
local connection 1-form of the connection $K_h$ on $L^hX$. We have
\ben
&&z^{h*}\om_L= \frac12 A_\la{}^{ab}dx^\la\ot e_{ab}, \label{K102} \\
&& A_\la{}^{ab} =\frac12 (\eta^{kb}h^a_\m-\eta^{ka}h^b_\m)(\dr_\la h^\m_k -
 h^\nu_k K_\la{}^\m{}_\nu). \nonumber 
\een
If $K$ is a Lorentz connection $A_h$, then obviously $K_h=A_h$.

The connection $K_h$  
(\ref{K102}) on $L^hX$ yields the corresponding spin connection on $S^h$
\beq
K_h=dx^\la\ot[\dr_\la +\frac14 (\eta^{kb}h^a_\m-\eta^{ka}h^b_\m)(\dr_\la
h^\m_k - h^\nu_k K_\la{}^\m{}_\nu)L_{ab}{}^A{}_B y^B\dr_A], \label{b3212}
\eeq
where $L_{ab}$ are the generators (\ref{b3213})  
(Giachetta and Mangiarotti, 1997; Sardanashvily, 1997). 
Such a connection has been considered by Ponomarev and Obuchov, 1982;
Aringazin and Michailov, 1991; Tucker and Wang, 1995.
Substituting the spin connection (\ref{b3212}) into the Dirac
operator (\ref{l13}) and the Dirac Lagrangian density (\ref{b3215}), we
obtain a description of Dirac fermion fields in the presence of an arbitrary
world connection, not only of the Lorentz
type.

One can use the connection (\ref{b3212}) to obtain a horizontal
lift onto $S^h$ of a vector field $\tau$ on $X$. This lift reads
\beq
\tau_{K_h} = \tau^\la\dr_\la +\frac14
\tau^\la(\eta^{kb}h^a_\m-\eta^{ka}h^b_\m) (\dr_\la
h^\m_k - h^\nu_k K_\la{}^\m{}_\nu) L_{ab}{}^A{}_B y^B\dr_A. \label{b3218}
\eeq
Moreover, we have the canonical horizontal lift
\beq
\wt\tau = \tau^\la\dr_\la +\frac14
(\eta^{kb}h^a_\m-\eta^{ka}h^b_\m) (\tau^\la\dr_\la
h^\m_k - h^\nu_k\dr_\nu\tau^\m) L_{ab}{}^A{}_B y^B\dr_A \label{b3216}
\eeq
of vector fields $\tau$ on $X$ onto the $h$-associated spinor bundle $S^h$
(Sardanashvily, 1997).
To construct the canonical lift (\ref{b3216}), one can write the canonical
lift of $\tau$ on the frame bundle $LX$ with respect to a Lorentz atlas
$\Psi^h$ and take its Lorentz part. Another way is the following. Let us
consider a local non-vanishing vector field $\tau$ and a 
local world symmetric
connection $K$ for which $\tau$ is an integral section. 
The horizontal lift (\ref{b3218}) of $\tau$ by means of
this connection is given by the expression (\ref{b3216}). 
In a straightforward manner, one can check that (\ref{b3216}) is a 
well-behaved lift of any vector
field $\tau$ on $X$.
The canonical lift (\ref{b3216}) is brought into the form
\be
\wt\tau = \tau_{\{\}} - \frac14
(\eta^{kb}h^a_\m-\eta^{ka}h^b_\m)h^\nu_k\nabla_\nu\tau^\m L_{ab}{}^A{}_B
y^B\dr_A,
\ee
where $\tau_{\{\}}$ is the horizontal lift (\ref{b3218}) of $\tau$ by means
of the spin Levi--Civita connection (\ref{b3217}) 
of the tetrad field $h$, and
$\nabla_\nu \tau^\m$ are the covariant derivatives of $\tau$ relative to the
same Levi--Civita connection (Kosmann, 1972; Fatibene {\it et al.}, 1996).

The canonical lift  (\ref{b3216}) fails to be a
generator of general covariant transformations because it does not involve
transformations of 
tetrad fields. To define general covariant transformations
of spinor bundles, we should consider spinor structures associated with
different tetrad fields. The
difficulty arises because, though the principal spinor bundles $P^h$ and
$P^{h'}$ are isomorphic, the associated structures of bundles of
Minkowski spaces $M^hX$ and $M^{h'}X$ (\ref{b3192}) on the cotangent
bundle $T^*X$ are not equivalent, and so are the
representations
$\g_h$ and $\g_{h'}$ (\ref{L4}) for different tetrad fields $h$ and $h'$ 
(Sardanashvily and Zakharov, 1992; Sardanashvily, 1995). 
Indeed, let
\be
t^*=t_\m dx^\m=t_ah^a=t'_a{h'}^a
\ee
be an element of $T^*X$. Its representations $\g_h$ and $\g_{h'}$ (\ref{L4})
read
\be
\g_h(t^*)=t_a\g^a=t_\m h^\m_a\g^a, \qquad 
\g_{h'}(t^*)=t'_a\g^a=t_\m {h'}^\m_a \g^a.
\ee
These representations are not equivalent because
no isomorphism $\Phi_s$ of $S^h$ onto $S^{h'}$ can obey the
condition
\be
\g_{h'}(t^*)=\Phi_s \g_h(t^*)\Phi_s^{-1}, \qquad \forall t^*\in T^*X.
\ee

It follows that every Dirac fermion field must be described 
by a pair with a
certain tetrad field.  We thus observe the phenomenon of
symmetry breaking in gauge gravitation theory which exhibits the physical
nature of gravity as a Higgs field (Sardanashvily, 1991; 
Sardanashvily and Zakharov, 1992). 

\section{Spontaneous symmetry breaking}

Spontaneous symmetry breaking is a quantum phenomenon 
modelled by classical Higgs fields.
In gauge theory on a principal bundle $P\to X$, the necessary condition for
spontaneous symmetry breaking to take place is the
reduction of the structure group
$G$ of $P$ to the subgroup $H$ of exact symmetries. Higgs
fields  are described by global sections
$h$ of the quotient bundle $\Si$ (\ref{b3193}). 

In accordance with Theorem \ref{redsub}, the set of Higgs fields $h$ is in
bijective correspondence with that of reduced $H$-principal subbundles
$P^h$ of $P$. Given such a subbundle $P^h$, let $Y^h$ (\ref{int1})
be the associated  bundle with a typical fibre $V$
which admits a representation of the group $H$ of exact symmetries, but
not the whole symmetry group $G$. Its sections $s_h$
describe matter fields in the presence of the Higgs fields $h$.
In general, $Y^h$ is not associated or 
canonically associated with other $H$-principal subbundles of $P$.
It follows that $V$-valued matter
fields can be represented only by pairs with
Higgs fields. The goal is to describe the totality of these pairs
$(s_h,h)$ for all Higgs fields (Sardanashvily, 1992, 1993).

Let us consider the composite bundle (\ref{b3223a}) and the composite
bundle 
\beq
Y\ar^{\pi_{Y\Si}} \Si\ar^{\pi_{\Si X}} X, \label{b3225}
\eeq
where $Y\to \Si$ is the bundle 
\beq
Y=(P\times V)/H \label{b3224}
\eeq 
associated with the $H$-principal bundle $P_\Si$ (\ref{b3194}).
There is the canonical isomorphism $i_h: Y^h \to h^*Y$
of the $P^h$-associated bundle 
$Y^h$ to the subbundle of $Y\to X$ which is the restriction
$h^*Y=(h^*P\times V)/H$ of the 
bundle $Y\to\Si$ to $h(X)\subset \Si$, i.e.,
\beq
i_h(Y^h)\cong \pi^{-1}_{Y\Si}(h(X)). \label{b3226}
\eeq
Then every global section $s_h$ of $Y^h$ corresponds to the
global section $i_h\circ s_h$ of the composite bundle (\ref{b3225}).
Conversely, every global section $s$ of the composite bundle
(\ref{b3225}) which projects onto a section $h=\pi_{Y\Si}\circ s$ of the
bundle $\Si\to X$ takes its values into the subbundle $i_h(Y^h)\subset
Y$ in accordance with the relation (\ref{b3226}). Hence, there is 
one-to-one correspondence between the sections of the bundle $Y^h$
and the sections of the composite bundle (\ref{b3225})
which cover $h$.

Thus, it is precisely the composite bundle 
(\ref{b3225}) whose sections describe
the above mentioned totality of the pairs $(s_h, h)$ of matter fields and
Higgs fields in gauge theory with broken symmetries (Sardanashvily and
Zakharov, 1992; Sardanashvily, 1991, 1995).

The feature of the dynamics of field systems on composite bundles
consists in the following.
Let the composite bundle $Y$ (\ref{b3225}) be coordinatized by $(x^\la,
\si^m, y^i)$, where $(x^\la, \si^m)$ are bundle coordinates on
$\Si\to X$.  Let 
\beq
A_\Si=dx^\la\ot(\dr_\la+ A^i_\la\dr_i)
+d\si^m\ot(\dr_m+A^i_m\dr_i) \label{b3228}
\eeq
be a principal connection on the bundle $Y\to \Si$.
This connection defines the splitting
\be
&& VY=VY_\Si\op\oplus_Y (Y\op\times_\Si V\Si), \\
&& \dot y^i\dr_i + \dot\si^m\dr_m=
(\dot y^i -A^i_m\dot\si^m)\dr_i + \dot\si^m(\dr_m+A^i_m\dr_i).
\ee
Using this splitting, one can construct
the first order differential operator
\ben
&&\wt D:J^1Y\to T^*X\op\ot_Y VY_\Si,\nonumber\\
&&\wt D=dx^\la\ot(y^i_\la- A^i_\la -A^i_m\si^m_\la)\dr_i,\label{7.10}
\een
on the composite bundle $Y$. 
The operator (\ref{7.10}) posesses the following property. 
Given a global section $h$ of $\Si$, its restriction 
\ben
&&\wt D_h =\wt D\circ J^1i_h: J^1Y^h \to T^*X\ot VY^h, \label{b3260}\\
&& \wt D_h =dx^\la\ot(y^i_\la- A^i_\la -A^i_m\dr_\la h^m)\dr_i, \nonumber
\een
to $Y^h$ is exactly the familiar covariant differential relative to the 
principal connection
\be
A_h=dx^\la\ot[\dr_\la+(A^i_m\dr_\la h^m + A^i_\la)\dr_i]
\ee
on the bundle $Y^h\to X$, which is induced by the principal connection
(\ref{b3228}) on the fibre bundle $Y\to \Si$ by the imbedding $i_h$
(Kobayashi and Nomizu, 1963).

\section{Universal spin structure}

All spin structures on a manifold $X$ which
are related to the two-fold universal covering groups possess the following
properties (Greub and Petry, 1978).
Let $P\to X$ be a principal bundle with a structure group $G$ with the
fundamental group $\pi_1(G)={\bf Z}_2$. Let $\wt G$ be the universal covering
group of $G$.
The topological obstruction to the existence of a $\wt G$-principal bundle
$\wt P\to X$ covering the bundle $P$ is given by the \v Cech
cohomology group
$H^2(X;{\bf Z}_2)$ of $X$ with coefficients in ${\bf Z}_2$. Roughly
speaking, the principal bundle $P$ defines an element of $H^2(X;{\bf
Z}_2)$ which must be zero so that $P\to X$ gives rise to $\wt P\to X$.
 Non-equivalent lifts of a $G$-principal bundle $P$ 
to a $\wt G$-principal bundle
are classified by elements of the \v Cech cohomology group 
$H^1(X;{\bf Z}_2)$.

In particular, the well-known topological obstruction to the existence of 
a spin structure
is the non-zero  second Stiefel--Whitney  class of $X$. 
In the case of 4-dimensional non-compact manifolds,
all pseudo-Riemannian spin structures are equivalent.

Let us turn to fermion fields in gauge gravitation theory, basing
our consideration on
the following two facts (Giachetta {et al.}, 1997).

\begin{prop}\label{ferm1}
The $L$-principal bundle 
\beq
P_L:=GL_4\to GL_4/L \label{b3244}
\eeq
is trivial.
\end{prop}

\begin{prop}\label{ferm2}
Since the first homotopy group of the group $GL_4$ is
${\bf Z}_2$, $GL_4$ admits the universal two-fold covering group
$\wt{GL}_4$. We have the commutative diagram
\beq
\begin{array}{ccc}
 \wt{GL}_4 & \longrightarrow &  GL_4 \\
 \put(0,-10){\vector(0,1){20}} & 
& \put(0,-10){\vector(0,1){20}}  \\
L_s & \ar^{z_L} &  L
\end{array} \label{b3243}
\eeq
\end{prop}

A  universal spin structure on $X$ is
defined to be a pair consisting of a
$\wt{GL}_4$-principal bundle
$\wt{LX}\to X$ and a principal bundle morphism over $X$
\beq
\wt z: \wt{LX} \to LX \label{b3247}
\eeq
(Dabrowski and Percacci, 1986; Percacci, 1986; Switt, 1993). 
This is unique since
$X$ is parallelizable. In virtue of Proposition
\ref{ferm2}, the diagram 
\beq
\begin{array}{ccc}
 \wt{LX} & \ar^{\wt z} &  LX \\
 \put(0,-10){\vector(0,1){20}} & 
& \put(0,-10){\vector(0,1){20}}  \\
P^h & \ar^{z_h} & L^hX 
\end{array} \label{b3222}
\eeq 
commutes for any tetrad field $h$ (Fulp {\it et al.}, 1994; Giachetta
{\it et al.}, 1997).
It follows that the quotient $\wt{LX}/L_s$ is
exactly the quotient $\Si_T$ (\ref{5.15}) so that there is the
commutative diagram
\beq
\begin{array}{rcl}
 \wt{LX}  & \op\longrightarrow^{\wt z} &  LX \\
  & \searrow  \swarrow & \\ 
 & {\Si_T} &  
\end{array} \label{b3250}
\eeq
Let us consider the composite bundle
\beq
\wt{LX}\to \Si_T \to X, \label{b3248}
\eeq
where $\wt{LX}\to \Si_T$ is the $L_s$-principal bundle. For each tetrad
field $h:X\to \Si_T$, the restriction of the $L_s$-principal bundle
$\wt{LX}\to \Si_T$ to $h(X)\subset \Si_T$ is isomorphic to the
$h$-associated principal spinor bundle $P^h$. Therefore, the
diagram (\ref{b3250}) is
called the  universal Dirac spin
structure.  

The universal Dirac spin structure (\ref{b3250}) can be regarded as
the
$L_s$-spin structure on the bundle of Minkowski spaces 
\be
E_M=(LX\times M)/L\to\Si_T
\ee
associated with the $L$-principal bundle
$LX\to\Si_T$.
Since the principal bundles  $LX$ and $P_L$ (\ref{b3244}) are trivial,
so is the
bundle $E_M\to \Si_T$. Hence, it is
isomorphic to the pull-back 
\beq
\Si_T\op\times_X T^*X. \label{b3252}
\eeq
One can show that a spin structure on this bundle is unique (Giachetta
{\it et al.}, 1997).

Let us consider the composite spinor bundle
\beq
S\ar^{\pi_{S\Si}}\Si_T\ar^{\pi_{\Si X}} X, \label{L1}
\eeq
where $S=(\wt{LX}\times V)/L_s$
is the spinor bundle $S\to \Si_T$ associated with the
$L_s$-principal bundle $\wt{LX}\to \Si_T$. 
Given a tetrad field $h$, there is the canonical isomorphism 
\be
i_h: S^h=(P^h\times V)/L_s \to (h^*\wt{LX}\times V)/L_s
\ee
of the $h$-associated spinor bundle $S^h$ (\ref{510}) onto the
restriction $h^*S$ of the spinor bundle $S\to \Si_T$ to $h(X)\subset
\Si_T$. Thence, every global section $s_h$ of the spinor bundle $S^h$
corresponds to the global section $i_h\circ s_h$ of the composite spinor 
bundle (\ref{L1}). Conversely, every global section $s$ of the composite
spinor bundle (\ref{L1}), which projects onto a tetrad field
$h$, takes its values into the
subbundle $i_h(S^h)\subset S$.

Let the frame bundle $LX\to X$ be provided with a holonomic atlas, 
and let the principal bundles $\wt{LX}\to \Si_T$ and
$LX\to\Si_T$ have the associated atlases $\{U_\e,z^s_\e\}$ and
$\{U_\e,z_\e=\wt z\circ z^s_\e\}$. With these atlases, the composite spinor
bundle
$S$ is equipped with the bundle coordinates $(x^\la,\si_a^\m, y^A)$, where
$(x^\la,
\si_a^\m)$ are coordinates on $\Si_T$ such that
$\si^\m_a$ are the matrix components of the group element
$(T\f_\zeta\circ z_\e)(\si),$
$\si\in U_\e,\, \pi_{\Si X}(\si)\in U_\zeta.$
For each section $h$ of $\Si_T$, we have
$(\si^\la_a\circ h)(x)= h^\la_a(x)$.

The composite spinor bundle $S$ is equipped with the fibre spinor metric
\be
a_S(v,v')=\frac12(v^+\g^0v' +{v'}^+\g^0v), \qquad 
\pi_{S\Si}(v)=\pi_{S\Si}(v').
\ee

Since the bundle of Minkowski spaces $E_M\to \Si_T$ is isomorphic
to the pull-back bundle (\ref{b3252}), there exists the representation 
\beq
\g_\Si: T^*X\op\ot_{\Si_T} S= (\wt{LX}\times (M\ot V))/L_s
\to (\wt{LX}\times\g(M\ot V))/L_s=S, \label{L7}
\eeq
given by the coordinate expression
\be
\wh dx^\la=\g_\Si (dx^\la) =\si^\la_a\g^a.
\ee
Restricted to $h(X)\subset \Si_T$, this representation  recovers the
morphism
$\g_h$ (\ref{L4}).

Using this representation, one can construct the total Dirac
operator on the composite spinor bundle $S$ as follows. 
Since the bundles which make 
up the composite bundle (\ref{b3248}) are trivial,
let us consider a principal connection $A_\Si$ (\ref{b3228}) on the
$L_s$-principal bundle $\wt{LX}\to
\Si_T$ given
by the local connection form
\beq
A_\Si = (A_\la{}^{ab} dx^\la+ A^k_\m{}^{ab} d\si^\m_k)\ot L_{ab},
\label{L10}
\eeq
where
\ben
&& A_\la{}^{ab} =-\frac12 (\eta^{kb}\si^a_\m-\eta^{ka}\si^b_\m)
 \si^\nu_k K_\la{}^\m{}_\nu, \nonumber\\
&& A^k_\m{}^{ab}=\frac12(\eta^{kb}\si^a_\m -\eta^{ka}\si^b_\m) \label{M4}
\een
and $K$ is a world connection on $X$. 
This connection  defines the associated spin connection
\beq
A_S = dx^\la\ot(\dr_\la + \frac12A_\la{}^{ab}L_{ab}{}^A{}_By^B\dr_A) +
d\si^\m_k\ot(\dr^k_\m +  \frac12A^k_\m{}^{ab}L_{ab}{}^A{}_By^B\dr_A)
\label{b3266}
\eeq
on the spinor bundle $S\to\Si_T$. Let $h$ be a global section
of $\Si_T\to X$ and $S^h$ the restriction of the bundle $S\to \Si_T$
to $h(X)$. It is readily observed that the restriction of the spin
connection (\ref{b3266}) to $S^h$ is exactly the spin connection
(\ref{b3212}).

The connection (\ref{b3266}) yields the first order differential
operator $\wt D$ (\ref{7.10}) on the composite spinor bundle $S\to X$ which
reads
\ben 
&&\wt D:J^1S\to T^*X\op\ot_{\Si_T} S,\nonumber\\
&&\wt D=dx^\la\ot[y^A_\la- \frac12(A_\la{}^{ab} + A^k_\m{}^{ab}\si_{\la
k}^\m)L_{ab}{}^A{}_By^B]\dr_A  =\label{7.10'} \\
&& \qquad dx^\la\ot[y^A_\la-
\frac14(\eta^{kb}\si^a_\m -\eta^{ka}\si^b_\m)(\si^\m_{\la k} -\si^\nu_k
K_\la{}^\m{}_\nu)L_{ab}{}^A{}_By^B]\dr_A. \nonumber
\een
The corresponding restriction $\wt D_h$ (\ref{b3260}) of the
operator $\wt D$ (\ref{7.10'}) to
$J^1S^h\subset J^1S$ recovers the familiar covariant differential on the
$h$-associated spinor bundle $S^h\to X$ relative to the spin connection
(\ref{b3216}).

Combining (\ref{L7}) and (\ref{7.10'}), we obtain 
the first order differential operator
\ben
&& \Delta=\g_\Si\circ\wt D:J^1S\to T^*X\op\ot_{\Si_T} S\to S,
\label{b3261}\\
&& y^B\circ\Delta=\si^\la_a\g^{aB}{}_A[y^A_\la-
\frac14(\eta^{kb}\si^a_\m -\eta^{ka}\si^b_\m)(\si^\m_{\la k} -\si^\nu_k
K_\la{}^\m{}_\nu)L_{ab}{}^A{}_By^B], \nonumber
\een
on the composite spinor bundle $S\to X$.
One can think of $\Delta$ as being the  total Dirac operator  on
$S$ since, for every tetrad field $h$, the restriction of $\Delta$ to
$J^1S^h\subset J^1S$  is exactly the 
Dirac operator $\Delta_h$ (\ref{l13}) on
the
$h$-associated spinor bundle
$S^h$
in the presence of the background tetrad field $h$ and the spin connection
(\ref{b3212}).

It follows that gauge
gravitation theory is reduced to the model of metric-affine
gravity and Dirac fermion fields.
The total configuration space of this model is the jet manifold
$J^1Y$ of the bundle product
\beq
Y=C_K
\op\times_{\Si_T} S \label{042}
\eeq
where 
$C_K$ is the bundle of world connections (\ref{015}). It is 
coordinatized by $(x^\m,\si^\m_a, k_\m{}^\al{}_\bt,y^A)$.

Let $J^1_\Si Y$ denotes the first order jet manifold of the bundle
$Y\to\Si_T$. This bundle can be provided with
the spin connection
\ben
&& A_Y: Y\ar J^1_\Si Y\ar^{\pr_2} J^1_\Si S, \nonumber\\
&&A_Y = dx^\la\ot(\dr_\la +\wt A_\la{}^{ab}L_{ab}{}^A{}_By^B\dr_A) +
d\si^\m_k\ot(\dr^k_\m +  A^k_\m{}^{ab}L_{ab}{}^A{}_By^B\dr_A), \label{b3263}
\een
where $A^k_\m{}^{ab}$ is given by the expression
(\ref{M4}), and
\be
\wt A_\la{}^{ab} =-\frac12 (\eta^{kb}\si^a_\m-\eta^{ka}\si^b_\m)
 \si^\nu_k k_\la{}^\m{}_\nu.
\ee
Using the connection (\ref{b3263}), we obtain the first order
differential operator
\ben 
&&\wt D_Y:J^1Y\to T^*X\op\ot_{\Si_T} S,\nonumber\\
&&\wt D_Y=dx^\la\ot[y^A_\la- 
\frac14(\eta^{kb}\si^a_\m -\eta^{ka}\si^b_\m)(\si^\m_{\la k} -\si^\nu_k
k_\la{}^\m{}_\nu)L_{ab}{}^A{}_By^B]\dr_A, \label{7.100}
\een
and the total Dirac operator
\ben
&& \Delta_Y=\g_\Si\circ\wt D:J^1Y\to T^*X\op\ot_{\Si_T} S\to S,
\label{b3264}\\
&& y^B\circ\Delta_Y=\si^\la_a\g^{aB}{}_A[y^A_\la-  \frac14(\eta^{kb}\si^a_\m
-\eta^{ka}\si^b_\m)(\si^\m_{\la k} -\si^\nu_k
k_\la{}^\m{}_\nu)L_{ab}{}^A{}_By^B],
\nonumber
\een
on  the bundle $Y\to X$. 

Given a section
$K:X\to C_K$, the restrictions of the spin connection $A_Y$ (\ref{b3263}),
the operator $\wt D_Y$ (\ref{7.100}) and the Dirac operator $\Delta_Y$
(\ref{b3264}) to $K^*Y$ are exactly the 
spin connection (\ref{b3266}) and the
operators (\ref{7.10'}) and (\ref{b3261}), respectively.

The total Lagrangian
density on the configuration space $J^1Y$ of the metric-affine gravity
and fermion fields is the sum $L=L_{MA} + L_D$
of the metric-affine Lagrangian density $L_{MA}$
and the Dirac Lagrangian
density
\ben
&& L_D=\{\frac{i}{2}\si^\la_q[y^+_A(\g^0\g^q)^A{}_B(y^B_\la-
\frac14(\eta^{kb}\si^a_\m
-\eta^{ka}\si^b_\m)(\si^\m_{\la k} -\si^\nu_k
k_\la{}^\m{}_\nu)L_{ab}{}^B{}_Cy^C)- \nonumber\\
&& \qquad (y^+_{\la A} -
\frac14(\eta^{kb}\si^a_\m
-\eta^{ka}\si^b_\m)(\si^\m_{\la k} -\si^\nu_k
k_\la{}^\m{}_\nu)y^+_C L^+_{ab}{}^C{}_A(\g^0\g^q)^A{}_By^B]- \label{b3265}\\ 
&&\qquad  my^+_A(\g^0)^A{}_By^B\}\sqrt{\nm\si}, \qquad \si=\det(\si_{\m\n}).
\nonumber
\een

It is readily observed that
\beq
\frac{\dr L_D}{\dr k_\la{}^\m{}_\nu} + 
\frac{\dr L_D}{\dr k_\nu{}^\m{}_\la} =0, \label{2C14}
\eeq
that is, the Dirac Lagrangian 
density (\ref{b3265}) depends only on the torsion
of a world connection. 

\section{General covariant transformations}

Since a world manifold $X$ is parallelizable and the 
universal spin structure
is unique, the
$\wt{GL}_4$-principal bundle
$\wt{LX}\to X$ as well as the frame bundle $LX$ admits the canonical lift of
any diffeomorphism $f$ of the base $X$. This lift is 
defined by the commutative
diagram
\be
\begin{array}{rcccl}
 &\wt{LX} & \ar^{\wt \Phi} & \wt{LX}& \\
_{\wt z} &\put(0,10){\vector(0,-1){20}} &  
& \put(0,10){\vector(0,-1){20}} &
_{\wt z} \\
& LX & \ar^{\Phi} & LX & \\
 &\put(0,10){\vector(0,-1){20}} &  & \put(0,10){\vector(0,-1){20}} & \\
& X & \ar^f & X  &
\end{array} 
\ee
where $\Phi$ is the 
holonomic bundle automorphism of $LX$ (\ref{025}) induced
by $f$ (Dabrowski and Percacci, 1986).

The associated morphism of the spinor bundle $S$ (\ref{L1}) is given by
the relation
\beq
\wt \Phi_S: (p, v)\cdot L_s \to (\wt \Phi(p), v)\cdot L_s, \qquad p\in
\wt{LX},\qquad v\in S. \label{b3270}
\eeq
Because $\wt \Phi$ is equivariant, this is a fibre-to-fibre automorphism
of the bundle $S\to \Si_T$ over the canonical automorphism of the
$LX$-associated tetrad bundle $\Si_T\to X$ induced by
the diffeomorphism $f$ of $X$. Thus, we have the commutative
diagram 
\be
\begin{array}{ccc}
 S & \ar^{\wt \Phi_S} & S \\
\put(0,10){\vector(0,-1){20}} &  & \put(0,10){\vector(0,-1){20}}  \\
\Si_T & \ar^{\Phi_\Si} & \Si_T \\
\put(0,10){\vector(0,-1){20}} &  & \put(0,10){\vector(0,-1){20}} \\
X & \ar^f & X  
\end{array} 
\ee
of general covariant transformations of the spinor bundle $S$.

Accordingly, there exists a canonical lift $\wt\tau$ onto $S$ 
of every vector field
$\tau$ on $X$. The goal is to find its coordinate expression. Difficulties
arise because the tetrad coordinates $\si^\m_a$ 
of $\Si_T$ depend
on the choice of an atlas of the bundle $LX\to \Si_T$. Therefore,
non-canonical vertical components appear in the coordinate expression of
$\wt\tau$.

A comparison with the canonical lift (\ref{973}) 
of a vector field $\tau$ over
the metric bundle $\Si_{PR}$ shows that the similar canonical lift of
$\tau$ over the tetrad bundle $\Si_T$ 
takes the form
\beq
\tau_\Si=\tau^\la\dr_\la + \dr_\nu\tau^\m \si^\nu_c \frac{\dr}{\dr \si^\m_c}
+Q^\m_c \frac{\dr}{\dr \si^\m_c}, \label{b3275}
\eeq
where the terms $Q^\m_c$ obey the condition
\be
(Q^\m_a\si^\nu_b + Q^\nu_a\si^\m_b)\eta^{ab}= 0.
\ee
The term
$Q^\m_a\dr_\m^a$ is the above-mentioned non-canonical part of the lift
(\ref{b3275}).

Let us consider 
a horizontal lift $\wt\tau_S$ of the vector field $\tau_\Si$
onto the spinor bundle $S\to \Si_T$ by means of the spin connection
(\ref{b3266}). It reads
\be
&& A_S\wt\tau_\Si=\tau^\la\dr_\la + \dr_\nu\tau^\m \si^\nu_c 
\frac{\dr}{\dr\si^\m_c}+ \\
&&
\qquad \frac14(\eta^{kb}\si^a_\m-\eta^{ka}\si^b_\m)
\si^\nu_k(\dr_\nu\tau^\m -K_\la{}^\m{}_\nu\tau^\nu) 
(L_{ab}{}^A{}_By^B\dr_A +L^+_{ab}{}^A{}_By^+_A\dr^B) +\\  
&& \qquad Q^\m_c \frac{\dr}{\dr \si^\m_c} +\frac14  Q^\m_k
(\eta^{kb}\si^a_\m-\eta^{ka}\si^b_\m)(L_{ab}{}^A{}_By^B\dr_A + 
L^+_{ab}{}^A{}_By^+_A\dr^B).
\ee 
Moreover, we obtain the desired canonical lift of $\tau$ onto $S$:
\ben
&&\wt\tau_S = \tau^\la\dr_\la + \dr_\nu\tau^\m \si^\nu_c 
\frac{\dr}{\dr\si^\m_c}+ \label{b3279}\\ 
&& \qquad Q^\m_c \frac{\dr}{\dr \si^\m_c} +\frac14  
Q^\m_k(\eta^{kb}\si^a_\m-\eta^{ka}\si^b_\m)
(L_{ab}{}^A{}_By^B\dr_A + L^+_{ab}{}^A{}_By^+_A\dr^B),\nonumber
\een
which can be written in the form
\be
&& \wt\tau_S = \tau^\la\dr_\la + \dr_\nu\tau^\m \si^\nu_c 
\frac{\dr}{\dr \si^\m_c} + \\
&& \qquad \frac14  Q^\m_k(\eta^{kb}\si^a_\m-\eta^{ka}\si^b_\m)
[-L_{ab}{}^d{}_c\si^\nu_d\frac{\dr}{\dr\si^\nu_c} +
L_{ab}{}^A{}_By^B\dr_A + L^+_{ab}{}^A{}_By^+_A\dr^B], 
\ee
where $L_{ab}{}^d{}_c$ are the generators (\ref{b3278}) (Giachetta {\it
et al.}, 1997). The corresponding
total vector field on the fibred product $Y$ (\ref{042}) reads
\ben
&& \wt\tau_Y = \wt\tau + \vt, \nonumber \\
&& \wt\tau =\tau^\la\dr_\la + \dr_\nu\tau^\m \si^\nu_c 
\frac{\dr}{\dr \si^\m_c} + \label{b3280'}\\
&&\qquad [\dr_\nu\tau^\al k_\m{}^\nu{}_\bt -\dr_\bt\tau^\nu 
k_\m{}^\al{}_\nu - \dr_\m\tau^\nu k_\nu{}^\al{}_\bt + \dr_{\m\bt}\tau^\al]
\frac{\dr}{\dr k_\m{}^\al{}_\bt}, \nonumber \\
&& \vt= \frac14  Q^\m_k(\eta^{kb}\si^a_\m-\eta^{ka}\si^b_\m)
[-L_{ab}{}^d{}_c\si^\nu_d\frac{\dr}{\dr\si^\nu_c} +
L_{ab}{}^A{}_By^B\dr_A + L^+_{ab}{}^A{}_By^+_A\dr^B]. \nonumber
\een
Its canonical part $\wt\tau$ (\ref{b3280'}) is the
generator of a local 1-parameter group of general covariant transformations
of the bundle $Y$, whereas the vertical vector field 
$\vt$ is the generator of a local 1-parameter group of principal (Lorentz)
automorphisms of the bundle $S\to\Si_T$. By construction, the total 
Lagrangian density $L$ obeys the relations
\ben
&& \bL_{J^1\vt}L_D=0. \label{K200'}\\
&& \bL_{J^1\wt\tau}L_{MA}=0, \qquad \bL_{J^1\wt\tau}L_D=0.\label{K200}
\een
The relation (\ref{K200'}) leads to the N\"other conservation law,
while the equalities (\ref{K200}) lead to the energy-momentum one 
(Giachetta and Mangiarotti, 1997; Sardanashvily, 1997).

\end{document}